# Data Modelling for the Evaluation of Virtualized Network Functions Resource Allocation Algorithms

Windhya Rankothge
Universitat Pompeu Fabra
windhya.rankothge@upf.edu

Franck Le
IBM Research
fle@us.ibm.com

Alessandra Russo
Imperial College
a.russo@imperial.ac.uk

Jorge Lobo
ICREA-Universitat Pompeu Fabra
jorge.lobo@upf.edu

To conduct a more realistic evaluations on resource allocation algorithms for Virtualized Network Functions (VNFs), researches need data on: (1) potential Network Functions (NFs) chains (policies), (2) traffic flows passing through these NFs chains, (3) how the dynamic traffic changes affect the NFs (scale out/in) and (4) different data center architectures for the cloud infrastructure. However, there are no publicly available real data sets on NF chains and traffic that pass through NF chains. Therefore, we have used data from previous empirical analyses [1], [2] and made some assumptions to derive the required data to evaluate resource allocation algorithms for VNFs. We developed four programs to model the gathered data and generate the required data. All gathered data and data modelling programs are publicly available at [3]. We have used these data for our work in [4] and [5]

## I. POLICY REQUESTS GENERATION

When generating policy requests for the NFC, the main factor to be considered is the type (e.g., small, medium, large size network) of the enterprise/user, that is requesting the policies. Depending on the type of the enterprise/user, the total number of NFs required, the number of NFs in a policy and types of the NFs in the policy can vary. The policies (chains of NFs) used in the experiments are generated based on a study about middle-boxes used in enterprise networks [1]. This data set from [1] includes figures about types of enterprise networks, number and types of middle-boxes used in these enterprise networks. According to [1], a chain of NFs consists of 2 to 7 NFs, mostly 2 to 5. So the number of NFs in a policy follows a truncated power-low distribution with exponent 2, minimum 2 and maximum 7.

According [1], as shown in Figure 1, large scaled enterprises, with 10k-100k hosts can have average: IP Firewalls: 46, Application firewalls: 9, WAN optimizers: 0, Proxies: 6, Gateways: 3, VPNs: 6, Load Balancers: 7, IDS/IPS: 23 and Total: 100.

### A. Policy requests generation program

We have considered large scaled enterprise networks where each network has 100 NFs. A chain of NFs consists of 2 to 7 NFs and the number of NFs in a policy follows a truncated power-low distribution with exponent 2, minimum 2 and maximum 7. The types of NFs in a policy are selected randomly, with different probabilities based on how many instances of each type of NFs can be there in the enterprise. Policy requests generation program is written in c++.

- Inputs to the program: number of large scaled enterprise networks
- Output of the program: a set of policies for each enterprise with 100 NFs

## II. INITIAL TRAFFIC DISTRIBUTION

After generating the policy requests, for simulating the traffic, we need traffic data where owners (enterprises/users) of the flows can be identified, so that we can differentiate the traffic passing through each policy. In the real-life situation, the clients traffic passing through the set of NFs will be directed to the different applications as web server, voip server etc according to the clients requests/needs. So the traffic load that each client is expecting can be different based on the applications client is handling [6]. For the experiments, we assume our clients are handling web based applications and the traffic used for the experiments is taken from a study about web traffic [2]. The data set includes HTTP traffic breakdown of 30,000 users for a day which is measured at three different vantage points of an Italian ISP. Traffic breakdown reports HTTP traffic for every 2 hours.

### A. Initial traffic distribution program

We use the traffic data for each enterprise given in [2] at the starting point of the HTTP traffic breakdown, and assume it as the initial total traffic flow that will pass through all the policy chains of the enterprise. We assume that the initial total traffic load is equally distributed over the set of policies of that enterprise. The initial traffic distribution program is written in c++.

- Inputs to the program: the set of policies, initial traffic load
- Output of the program: distribution of the traffic load over policies

## III. SCALING REQUIREMENTS OVER THE TIME

In a data center, traffic changes happen throughout the day and according to the amount of these changes, the NFs should be scaled out/in to satisfy the dynamic demands. The limitation of our data set is it lacks the information on how the traffic changes occurred over two hours. It has information only on traffic at each two hours.

According to [7], as shown in Figure 2, traffic changes on usual days happen gradually over time. Even at events where traffic will be increased in a huge amount (elephant flows), as shown in Figure 3, the change happens gradually over a 15 minutes time period [8]. So we have assumed that, for the every 2 hours traffic reported in [2], increase/decrease happened uniformly through 2 hours and generated the traffic graph in Figure 4. It shows traffic flow for 24 hours in 10MBps units for each enterprise.

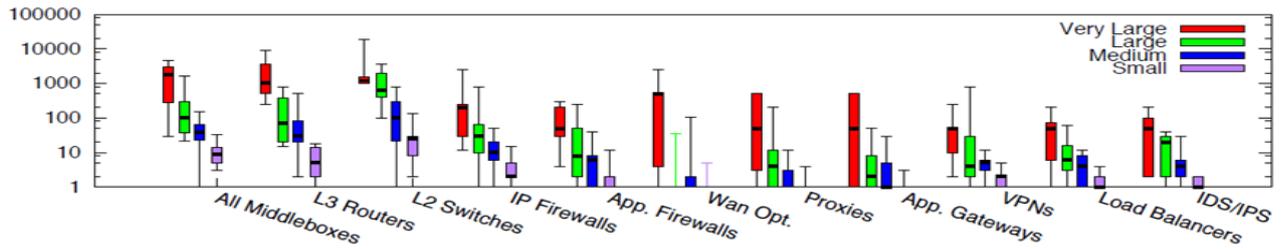

**Fig. 1:** Box plot of middlebox deployments for small (fewer than 1k hosts), medium (1k-10k hosts), large (10k-100k hosts), and very large (more than 100k hosts) enterprise networks. Y-axis is in log scale. [1]

But in special situations, there can be flash events, where sudden traffic changes occurred within few minutes. So we have to consider two situations: usual traffic patterns and elephant flows where traffic changes gradually and flash events where traffic changes suddenly. To reflect scaling requirements of both situations, we spread the increase/decrease of number of NFs (needed for the full 2 hours traffic change) over 2 hours and add/remove one instance at a time.

According to [9], as shown in Figure 5, if we add more than one instance at a time, to be ready for the requirements in the future, we are adding more than what is needed and wasting resources. So we define a threshold (Maximum amount of traffic that an instance of a NF can handle) to find how many instances we should add/remove to accommodate traffic change and we will add/remove one instance at a time.

### A. Scaling requirements over the time program

In the scaling requirements over the time program, first we define the threshold L, The maximum amount of traffic that an instance of a NF can handle. If the traffic change of a 2 hours period is greater than L, we assume that we have to add an extra instance of the NF. We are making an assumption: the traffic flowing through the NF instance is proportionate to the capacity of the NF instance and it is same for all types of NFs. In reality this might not be correct and different NF types may have different connections between traffic flow and capacity.

As the second step, we identify the enterprises whose traffic has changed over each 2 hours from the traffic graph. For each enterprise, we have already generated x number of policies, and assume each policy has a unique traffic flow passing through. When there is a change in the total traffic for that enterprise, it is very unlikely that traffic passing through all the policies of that enterprise contributed to the traffic change. Most probably the traffic change was caused by the traffic passing through sub set of policies. So for enterprises that have a traffic change, we randomly select 5 of its policies, as the policies affected by the traffic change.

After selecting the policies affected by each enterprise traffic change, the third step is to identify which NF from each policy, needs to be scaled out/in to satisfy the new traffic demands. According to Stratos [10], there are simple approaches we could leverage for deciding which NF(s) to scale. The simplest solution is to scale all NFs in that policy. This guarantees that any bottleneck will be eliminated, but this potentially wastes significant resources and imposes unneeded costs, when only one NF may be the bottleneck. So Stratos performs a set of scaling trials, scaling each NF in the policy, one (VM) instance at a time. They begin by adding a new instance of the first NF in the chain, monitoring for changes over a fixed time window. If performance improves beyond some threshold, then the new instances is permanently added to the tenants topology. No improvement means that the NF is not a bottleneck, so they discard the new instance. Then move to the next NF in the chain and repeat the process. Their results show that no two NFs will be simultaneously and equally bottlenecked and scaling one NF in the policy at a time is acceptable. Hence assuming the conditions in Stratos, we randomly select a NF from each policy as the bottlenecked NF for which the resource allocation needs to be increase/decrease.

The last step is to decide, from the identified NF instance to scale, how many instances we should add/remove to satisfy the new traffic demand. For each enterprise whose traffic has changed, first we identify the total traffic change over each 2 hours: $C$ from the traffic graph. Then we calculate how many instances had to be add/remove for each enterprise: $I$ based on the threshold $L$ we defined earlier.

$$I = C/L$$

If we have to add/remove instances, we spread the $I$ over 2 hours (120 minutes). As explained earlier, we are trying to add/remove instance at a time. Therefore, If I = 2, and starting time of the period is T=0, then scaling occurs when T+40 minutes and T+80 minutes. If we dont have to add/remove instances, we have to change the paths of the policies which use overloaded links because of traffic change.

Following the above process, the scaling requirements over the time program is written in c++.

- Inputs to the program: the set of policies, traffic pattern
- Output of the program: a set of policies and NFs effected by traffic changes during each interval and the required add/remove NF instances for each interval

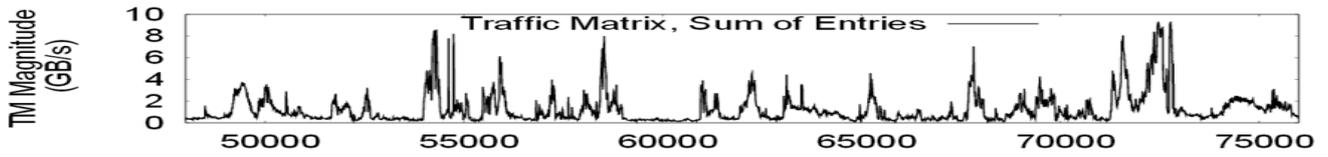

**Fig. 2:** Traffic in the data-center changes in the magnitude (Time in seconds). [7]

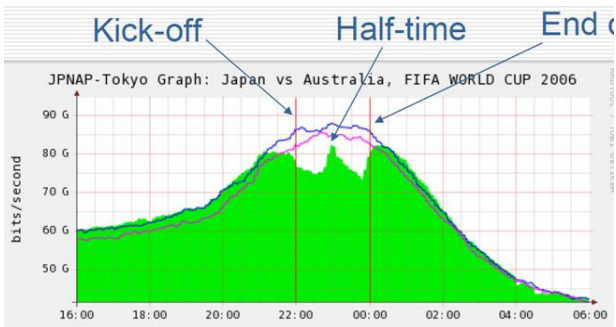

**Fig. 3:** Traffic statistics from World Cup 2006 [8]

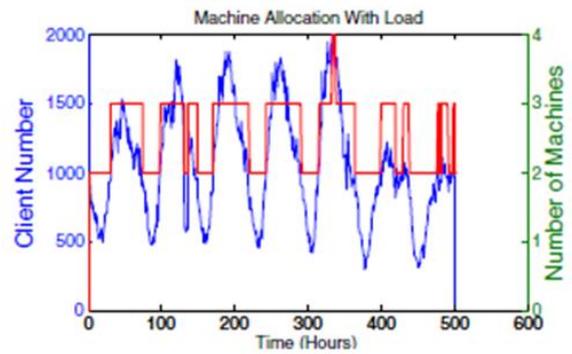

**Fig. 5:** Machines Allocation [9]

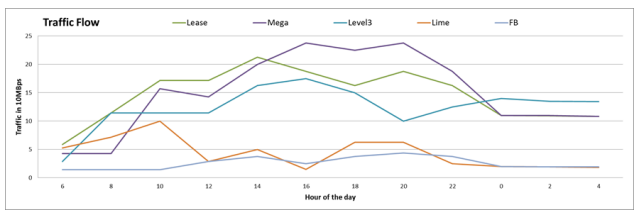

**Fig. 4:** Traffic over a full day

## IV. TOPOLOGY GENERATION

For the cloud network infrastructure, we have assumed three different data center network architectures: (1) k fat tree, (2) VL2 and (3) BCube as shown in figure 6. For each network architecture, we needed data on: (1) nodes of the network (servers and switches of the network), (2) links of the network (connecting two nodes), and (3) paths of the network (between each and every server of the network). All these three depends on two factors: (1) the number of servers in the cloud infrastructure and (2) network architecture of the cloud infrastructure.

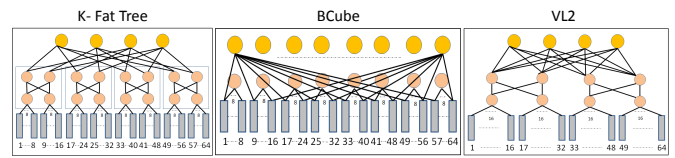

**Fig. 6:** (1) k fat tree, (2) BCube and (3) VL2

### A. Topology generation program

Following the standards given for (1) k fat tree, (2) VL2 and (3) BCube architectures [11], [12], [13], we define the network and equally distribute the number of servers over ToR switches of the network. The topology generation program is written in python.

- Inputs to the program: network architecture and number of servers

- Output of the program: the topology: nodes, links and paths

ACKNOWLEDGMENT

This research was sponsored by the U.S. Army Research Laboratory and the U.K. Ministry of Defence and was accomplished under Agreement Number W911NF-06-3-0001. The views and conclusions contained in this document are those of the author(s) and should not be interpreted as representing the official policies, either expressed or implied, of the U.S. Army Research Laboratory, the U.S. Government, the U.K. Ministry of Defence or the U.K. Government. The U.S. and U.K. Governments are authorized to reproduce and distribute reprints for Government purposes notwithstanding any copyright notation here on. Also this work is supported by the Maria de Maeztu Units of Excellence Programme (MDM-2015-0502).